# Theoretical Formulation of the Origin of cataclysmic Late Heavy Bombardment Era based on the New Perspective of Birth & Evolution of Solar Systems.

### 1. INTRODUCTION

In recent days four observations strongly suggest that in remote past Jupiter and the gas giants may have experienced gravitational sling shot and they may have been launched on an outward spiral path just the way Moon has been launched [personal communication: http://arXiv.org/abs/0805.0100 ] or for that matter all planetary natural satellites have been launched.

(A) 700 Hilda asteroids in elliptical orbit [Franklin et al 2004].The asteroid belt is populated with hundred thousands of rocky remnants leftover from planet formation. These are called asteroids and they lie between Mars and Jupiter orbit between a radii of 3AU to 10AU. Most of the asteroids are in near circular orbits. There are 700 odd asteroids known as Hilda which are in highly elliptical orbit and these eccentricities could have been imparted only by a migrating Jupiter set on an expanding spiral path. The migrating Jupiter first ejected some proto-Hilda asteroids out of the system and next elongated the orbits of the residual asteroids. The migrating Jupiter could have also set the planetary embryos on unruly chaotic paths which led to infrequent collisions and accretion resulting into terrestrial rocky planets.

(B) Through computer simulation studies [Tsiganis, Gomes, Morbidelli & Lavison 2005] it has been shown that our planetary system, with initial quasi-circular, coplanar orbits, would have evolved to the current orbital configurations provided Jupiter and Saturn crossed the 1:2 mean motion resonance (MMR). When the ratio of the orbital periods of Jupiter and Saturn is 1:2 it is the strongest resonance point. At all integer ratios resonance is obtained but the maximum is obtained at 1:2. The resonance crossings excite the orbital eccentricities and mutual orbital inclinations to the present values. Jupiter ,Saturn and Uranus have the present eccentricities of 6%, 9% and 8% respectively . The present mutual inclination of the orbital planes of Saturn, Uranus and Neptune take the maximum values of approximately 2º with respect to that of Jupiter. The simulation was started with the initial positions of Jupiter and Saturn at 5.45AU and 8AU respectively. 1:2MMR crossing occurs at 8.65AU. The present orbital semi major axes of Jupiter, Saturn, Uranus and Neptune are 10AU, 15AU, 19.3AU and 30AU respectively. This simulation reproduces all aspects of the orbits of the giant planets: existence of natural satellites, distribution of Jupiter's Trojans and the presence of main belt asteroids.

(C) The presence of Jupiter's Trojans can be explained only by 1:2MMR crossing by Jupiter and Saturn[Morbidelli, Levison, Tsiganis and Gomes 2005]. These are asteroids which are in he same orbit as that of Jupiter but they are leading or lagging by 60º in their co-orbital motion.

(D) The petrology record on our Moon suggests that a cataclysmic spike in the cratering rate occurred approximately 700 million years after the planets formed[Gomes, Levison, Tsiganis and Morbidelli 2005]. With the present evidence we assume the birth of our Solar Nebula at 4.56Gya. The formation of Gas Giants and Ice Giants was completed in first 5 millon years and Earth was completed in first 30 million years. This puts the date of completion of Giant Planets at 4.555Gya and the date of completion of the Terrestrial Planets particularly Earth at 30 million years after the solar nebula was born that is at 4.53Gya. At 4.53Gya, the Giant Impact occurred and from the impact generated circumterrestrial debris, Moon was born beyond Roche's Limit at 16,000Km orbital radius. By gravitational sling shot effect it was launched on an outward spiral path. Presently Moon is at the semi-major axis of 3,84,400Km with a recession velocity of 3.7cm/year. Towards the end of planet formation phase, the residual debris of the solar nebula was being rapidly sucked in or swept out of the system. This resulted in heavy meteoritic bombardment of all the big sub-stellar objects including our Moon. Through Apollo Mission studies it has been determined that there is a sharp increase in the bombardment rate and hence in the cratering rate around the period of 4.5 to 3.855Gya. From this it is concluded that there was a cataclysmic Late Heavy Bombardment of all big sub-stellar bodies, including our Moon, at about 700 My after the completion of formation of Jupiter and Saturn.

As a result of this outward migration considerable eccentricities were imparted to 700 odd Hilda asteroids [Franklin et al 2004]. Further more through simulation study it is concluded that during the course of orbital evolution when 1:2 MMR crossing is done the stimulus for Late Heavy Bombardment Era(LHB Era) is triggered [Gomes et al 2005]. We know that LHB Era lasted from 4Gya to 3.8Gya. Hence any satisfactory and adequate theory of migration of Giant Planets must explain this particular date of triggering of LHB Era. Meaning by if indeed the New Perspective on the birth and evolution of Solar System [Sharma & Ishwar 2004a,2004b] is a realistic picture of two body problem then we should be able to show the 1:2MMR crossing in this time window of time line of Birth and Evolution of our Solar System.

## 2. METHODOLOGY.

The assumed time line of birth and evolution of our Solar System is given in Table 1.

**Table 1. The Timeline of Planetary Formation[Santos et al 2005].**

| Triggering | Birth of Solar Nebula | Dissipation of gas & dust disk | Last Giant Impact | Late Heavy Bombardment | Life |
|---|---|---|---|---|---|
| 4.568Gya | 4.567Gya | 4.558Gya | 4.468Gya | 4.0Gya | 3.568Gya |
|  | 0 | 9My | 99My | 567My | 999My |
| Triggering | A supernova explosion in our neighborhood generates shock waves which sets a passing-by interstellar cloud of gas and dust into spin mode. This spinning primordial cloud flattens into a pancake like disc of cloud and dust. | | | | |
| Birth of Solar Nebula | The dust particles may be colliding and sticking giving rise to pebble sized solids. These pebbles further coalesce to form km-size planetesimals. This formally marks the birth of Solar Nebula and the start of Planetary Formation. Planetesimals collide and accrete to form planetary embryos. | | | | |

| Dissipation of gas and dust disk. | Particles less than micron size and gases are pushed out by photon pressures. This is known as Photo-evaporation. Particles of micron size and more are acted upon by Robertson-Poynting drag which constrains these particles to spiral inward and eventually fall into the host body. This leads to gradual dispersal and dissipation of gas-dust disk. By this process all the gas and dust will be removed in 30My. This means that within this narrow time slot the Gas Giants should have completed their formation. Hence in first 30My Jupiter and Saturn should complete their formation. Planetary embryos get enveloped by Hydrogen gas through gravitational accretional runaway mechanism terminated by the paucity of material because of a gaping void. When the void gets filled up then the next sequential gravitational accretional runaway process initiated. This process is repeated until all the gases are exhausted. In this way in 30My Jupiter, Saturn, Neptune and Uranus formation is completed. |
|---|---|
| Last Giant Impact | The terrestrial planets are not formed by runaway gravitational accretional mechanisms because such large amounts of material is not present to sustain such a process. Instead a series of infrequent and titanic impacts caused the formation of the present sized Earth, Venus, Mars and Mercury. The Giant Impact was the last such event, atleast in the context of Earth, which formally marked the completion of formation of Earth. |
| Late Heavy Bombardment Era. | 1:2 MMR crossing occurred by the spirally expanding orbits of Jupiter and Saturn. This triggered Neptune to be flung into Oort's Cloud. The disturbance of Oprt's caused a large amount of comets and asteroids to be flung into the inner part of the Solar System. This caused all the planets to experience a Late Heavy Bombardment Era about 567My after the birth of Solar Nebula,. The foot prints of this era is well preserved in the petrological record of Moon. |
| Life | After 1 Gy the first organic life based on anaerobic fermentation was initiated. |

It is assumed that Solar Nebula is born 4.567Gy Before the Present. While the central part is collapsing into Sun simultaneously Jupiter is born. It is assumed that Jupiter is fully formed in 7My leaving a gaping hole. That is the age of Jupiter is 4.56Gy. This paucity of raw material is what terminates the gravitational runaway accretion of Jupiter. Jupiter, due to gravitational sling shot effect, is launched on an expanding spiral path. As Jupiter rapidly spirals out, hydrodynamic stability is restored and the gaping hole in the disc of accretion is filled up with raw material for newer planet Saturn formation. Saturn could have formed 10My after the birth of Solar Nebula or 15My or 20My. The orbital path of both Jupiter and Saturn are plotted for first 1000My and the evolving semi major axis $a_S$ and $a_J$ for Saturn and Jupiter respectively are tabulated and from $(a_S/a_J)^{3/2}$ the Mean Motion Resonance (2/1) is examined. The cubic power of the square root of $(a_S/a_J)$ is the ratio of $P_S$ ( orbital period of Saturn) and $P_J$ ( orbital period of Jupiter).

2:1MMR crossing is examined for simultaneous birth of Jupiter and Saturn, for time difference of 5My, 10My, 15My, 20My, 25My with Jupiter preceding Saturn.

Saturn cannot be formed any later because in about 30My the protoplanetary gas-dust disk is completely dissipated by Poynting-Robertson Photon Drag and by Photoevaporation. In 30 My from the time of Solar Nebula all the four Jovian Planets formation should be completed. That is Jupiter, Saturn , Neptune and Uranus birth and formation should be completed sequentially in the descending order of their mass. As the time difference between the birth of Jupiter and Saturn widens, Jupiter birth-date is not pushed back but Saturn's birth date is advanced but not beyond 25My after the birth of Solar Nebula. Keeping this time constraint in mind the possible set of the birth dates are tabulated in Table 1 . This set of birth dates are taken while examining the various scenarios of birth and evolution of Jupiter ands Saturn.

**Table 1. Set of physically tenable birth dates of Jupiter and Saturn.**

| Jupiter Birth date(Gya) | Saturn Birth date(Gya) | Difference in The birth dates(My) |
|---|---|---|
| 4.56 | 4.560 | 0 |
| 4.56 | 4.555 | 5 |
| 4.56 | 4.550 | 10 |
| 4.56 | 4.545 | 15 |
| 4.56 | 4.540 | 20 |
| 4.555 | 4.535 | 25 |

The trajectories of Jupiter and Saturn are worked out for this set of birth dates and 2:1MMR crossing point examined. The crossing points are tabulated in Table 11 .

### 3. JUPITER'S ORBITAL EVOLUTION.

The Globe parameter of Sun is given in Table 2.

**Table 2. Globe Parameters of our Sun. [Chaisson et al 1998, Hannu et al 2003, Moore 2002]**

|  | M (kg) | R (m) | C (kg-m$^2$) | Period | $J_{spin}$ (kg-m$^2$-sec$^{-1}$) |
|---|---|---|---|---|---|
| Sun | $1.99 \times 10^{30}$ | $6.96 \times 10^8$ | $3.856 \times 10^{47}$ | 24.9d | $1.126155405 \times 10^{42}$ |

The Globe and Orbit parameters of Jupiter and Saturn are given in Table 3.

**Table 3. The Globe and Orbit parameters of Jupiter and Saturn [Chaisson et al 1998, Hannu et al 2003, Moore 2002]**

|  | $a$ ($\times 10^9 m$) | α (inclination angle in degress) | Φ (obliquity angle in degrees) | $P_1$ (d) | $P_3$ (d) | m (kg) | ρ + ($Kg/m^3$) |
|---|---|---|---|---|---|---|---|
| **Jupiter** | 778.3 | 1.31 | 3.1 | 4,331.865 | 0.41 | $1.90 \times 10^{27}$ | 1330 |
| **Saturn** | 1,427 | 2.49 | 26.7 | 10,760.265 | 0.43 | $5.69 \times 10^{26}$ | 710 |

$P_1$= orbital period of the Planet.
$P_3$= spin period of the Planet.

**Table 4. Tabulation of ω/Ω Equation Parameters for the Jupiter & Saturn($R_+$=Radius of the Planet & $J_{spin}|_+$=Spin Angular Momentum of the Planet, $J_T = (J_{spin})_\Theta + J_{orb} + J_{spin}|_+$ )**

|  | B (×$10^{10}$m$^{3/2}$/s) | $R_+$ (×$10^6$ m) | $J_{spin}|_+$ (Kg-m²/sec) | $J_{orb}$ (Kg-m²/sec) | $J_T$ (×$10^{42}$ Kg-m²/sec) | E (×$10^{-16}$ m$^{-3/2}$) | F (m$^{-2}$) |
|---|---|---|---|---|---|---|---|
| Jupiter | 1.152647951 | 71.492 | 6.89^38 | 1.93^43 | 20.42684441 | 45.95926 | 4.9227^-21 |
| Saturn | 1.152262984 | 60.268 | 1.398^38 | 7.83^42 | 8.956295205 | 20.15789 | 1.475^-21 |

**Table 5. Comparative tabulation of Jupiter and Saturn masses (m), Roche's Limits ($a_R$), Inner ($a_{G1}$) & Outer Geo-Synchronous Orbits ($a_{G2}$), the Evolution Factor [ Є= (a- $a_{G1}$)/( $a_{G2}$ - $a_{G1}$ ) ], calculated and observed ω/Ω.**

|  | a (×$10^9$m) | $a_R$ (×$10^9$m) | $a_{G1}$ (×$10^9$m) | $a_{G2}$ (×$10^{11}$m) | m (kg) | Є | ω/Ω\|calculated | ω/Ω\|observed |
|---|---|---|---|---|---|---|---|---|
| Jupiter | 778.3 | 1.7346 | 3.78579 | 8.71161 | 1.90×$10^{27}$ | 0.893 | 173.726 | 173.96 |
| Saturn | 1,427 | 2.138 | 6.52651 | 18.6641 | 5.69×$10^{26}$ | 0.764 | 432.189 | 432.139 |

As given in my personal communication [ http://arXiv.org/abs/0805.1454 ] about the evolutionary history of Mars satellite Phobos, in exactly the same manner the radial velocity expression is set up and the evolution of the semi-major axes of Jupiter and Saturn are calculated for the six sets of physically tenable birth dates as given in Table 1.

For calculating the Jupiter's Orbital Evolution, we calculate the geosynchronous orbits and gravitational resonance point. Next we calculate the structure factors in the expression for radial velocity and we check if the integration of the reciprocal of radial velocity from $a_{G1}$ to present semi-major axis gives the transit time of 4.56Gy as assumed in this paper.

The ω/Ω equation, which is equivalent to lom/lod in planetary satellite dynamics, is set up:

$$E \times x^{1.5} - F \times x^2 \;/.\; \{E \to 45.95926 \times 10^{-16}, F \to 4.9227 \times 10^{-21}\}$$

$$4.59592600000000039\text{`}^\wedge\text{-}15 \; x^{1.5} - 4.92269999999999985\text{`}^\wedge\text{-}21 \; x^2$$
8.1

Eq. 8.1 is ω/Ω equation and it is tested at the present semi-major axis to see if the calculated value of ω/Ω corresponds to the observed value.

$$4.59592600000000039\text{`*^-15 } x^{1.5`} - 4.92269999999999985\text{`*^-21 } x^2 \text{ /. } x \to 778.3 \times 10^9$$

$173.754567611763377\text{`} = $ calculated value of $\omega/\Omega$;

The observed value of $\omega/\Omega = 173.96$. Hence Eq.8.1 is correctly set up.

Geo-synchronous orbits radius is calculated by equating Eq.8.1 to UNITY.

$$\text{Solve}[4.59592600000000039\text{`*^-15 } x^{1.5`} - 4.92269999999999985\text{`*^-21 } x^2 == 1, x]$$

$\{\{x \to 3.78579271883727841\text{`*^9}\}, \{x \to 8.71177921068034066\text{`*^11}\}\}$

The above two values are $a_{G1}$ and $a_{G2}$.

Gravitational Resonance point is calculated by equating Eq.8.1 to TWO.

$$\text{Solve}[\\ 4.59592600000000039\text{`*^-15 } x^{1.5`} - 4.92269999999999985\text{`*^-21 } x^2 == 2,\\ x]$$

$\{\{x \to 6.08652819331731542\text{`*^9}\}, \{x \to 8.70710874762818143\text{`*^11}\}\}$

Hence $x_2$ ( the point of velocity maxima) is $6.0865 \times 10^9$ m.

Now the exponent M of the structure factor is calculated by equating the first time derivative of radial velocity to ZERO at $x_2$.

$$(-0.5 + M)/x^{0.5} + E \times (2 - M) x - F \times (2.5 - M) x^{1.5} \text{ /. } \{E \to 45.95926 \times 10^{-16}, F \to 4.9227 \times 10^{-21}\}$$

First E and F numerical values are substituted in the first time derivative of radial velocity. The result is:

$$\frac{-0.5\text{`} + M}{x^{0.5\text{`}}} + 4.59592600000000039\text{`*^-15 } (2 - M) x - 4.92269999999999985\text{`*^-21 } (2.5\text{`} - M) x^{1.5\text{`}}$$

Now $x_2$ numerical value is substituted to obtain the first derivative at velocity maxima point:

$$\frac{-0.5\text{`} + M}{x^{0.5\text{`}}} + 4.59592600000000039\text{`*^-15 } (2 - M) x - 4.92269999999999985\text{`*^-21 } (2.5\text{`} - M) x^{1.5\text{`}} \text{ /. } x \to 6.08652819331731542\text{`*^9}$$

The final first derivative at $x_2$ is:

$$0.0000279732331734000849\grave{\ } (2 - M) - 2.3375334095196858\grave{\ }{}^{\wedge}\text{-}6\, (2.5\grave{\ } - M) + 0.0000128178498819401909\grave{\ } (-0.5\grave{\ } + M)$$

8.2

Eq.8.2 is equated to ZERO to obtain the exponent M:

```
Solve[0.0000279732331734000849` (2 – M) –
    2.3375334095196858`*^-6 (2.5` – M) +
    0.0000128178498819401909` (–0.5` + M) == 0,
 M]
```

The root is:

M= {{M → 3.40881725753344833`}}

Now constant K is determined. Through several iterations it is found that $V_{max}$= 203017m/year gives a transit time of 4.56Gy from $a_{G1}$ to present semi-major axis.

Following Mathematica Commands are executed to determine K.
The radial velocity expression is first expressed in terms of m/year by multiplying the m/sec expression by $31.5569088 \times 10^6$ seconds/solar year and substituting the numerical values of E and F the constants of ω/Ω Equation.

$$(K \div x^M) \times (E \times x^2 - F \times x^{2.5} - \sqrt{x}) \times (2 \times (1 + m/M) \div (mB)) \times 31.5569088 \times 10^6 \;/.\; \{E \to 45.95926 \times 10^{-16}, F \to 4.9227 \times 10^{-21}\}$$

The result is:

$$\frac{1}{mB}\left(6.31138175999999884\grave{\ }{}^{\wedge}7\, K \left(1 + \frac{m}{M}\right) x^{-M} \left(-\sqrt{x} + 4.59592600000000039\grave{\ }{}^{\wedge}\text{-}15\, x^2 - 4.92269999999999985\grave{\ }{}^{\wedge}\text{-}21\, x^{2.5}\right)\right)$$

Next mass of Sun $m_0$ and mass of planet Jupiter are substituted the radial velocity expression:

$$\frac{1}{m_1 B}\left(6.31138175999999884\grave{\ }{}^{\wedge}7\, K\, (1 + m_1/m_0)\, x^{-M} \left(-\sqrt{x} + 4.59592600000000039\grave{\ }{}^{\wedge}\text{-}15\, x^2 - 4.92269999999999985\grave{\ }{}^{\wedge}\text{-}21\, x^{2.5}\right)\right) /.$$
$$\{m_1 \to 1.9 \times 10^{27}, m_0 \to 1.99 \times 10^{30}\}$$

The result is:

$$\frac{1}{B}\left(3.32495142230732554\grave{}\text{^}\text{-}20\,K\,x^{-M}\left(-\sqrt{x} + 4.59592600000000039\grave{}\text{^}\text{-}15\,x^2 - 4.92270000000000074\grave{}\text{^}\text{-}21\,x^{2.5}\right)\right)$$

Next numerical value of B= √[G(M+m)] is substituted in radial velocity expression:

$$\frac{1}{B}\left(3.32495142230732554\grave{}\text{^}\text{-}20\right.$$
$$\left. K\,x^{-M}\left(-\sqrt{x} + 4.59592600000000039\grave{}\text{^}\text{-}15\,x^2 - \right.\right.$$
$$\left.\left.4.92270000000000074\grave{}\text{^}\text{-}21\,x^{2.5}\right)\right)\,/.$$
$$B \to 1.152647951 \times 10^{10}$$

The result is:
$$2.88462007798886475\grave{}\text{^}\text{-}30\,K\,x^{-M}\left(-\sqrt{x} + 4.59592600000000039\grave{}\text{^}\text{-}15\,x^2 - 4.92270000000000074\grave{}\text{^}\text{-}21\,x^{2.5}\right)$$

Next the numerical value of exponent M is substituted:

$$2.88462007798886475\grave{}\text{^}\text{-}30$$
$$K\,x^{-M}\left(-\sqrt{x} + 4.59592600000000039\grave{}\text{^}\text{-}15\,x^2 - 4.92270000000000074\grave{}\text{^}\text{-}21\,x^{2.5}\right)\,/.$$
$$M \to 3.40882$$

The result is:

$$\left(2.88462007798886368\grave{}\text{^}\text{-}30\,K\left(-\sqrt{x} + 4.59592600000000039\grave{}\text{^}\text{-}15\,x^2 - 4.92270000000000074\grave{}\text{^}\text{-}21\,x^{2.5}\right)\right) / x^{3.40881999999999996\grave{}}$$

Now we have the radial velocity expression with only K unknown. To determine K we calculate the expression at $x=x_2$:

$$\left(2.88462007798886368\grave{}\text{^}\text{-}30\,K\left(-\sqrt{x} + 4.59592600000000039\grave{}\text{^}\text{-}15\,x^2 - 4.92270000000000074\grave{}\text{^}\text{-}21\,x^{2.5}\right)\right) /$$
$$x^{3.40881999999999996\grave{}}\,/.$$
$$x \to 6.08652819331731542\grave{}\text{^}9$$

The result is $V_{max}$ expression with only K unknown:

$$9.97972402108637979\grave{}\text{^}\text{-}59\,K$$

As already discussed that by several iterations we have found that $V_{max}$ should be 203017m/year in order to achieve an Age of 4.56Gy for Jupiter. Therefore the above expression is equated to 203017m/yr:

`Solve[9.979724021086379796`^-59 K == 203017, K]`

The unknown K is:

`{{K → 2.03429473170842056`^63}}`

By substituting the numerical value of K we set up the complete radial velocity expression:

$$\left(2.88462007798886368{}^{\wedge}\text{-}30\, K \left(-\sqrt{x} + 4.59592600000000039{}^{\wedge}\text{-}15\, x^2 - 4.92270000000000074{}^{\wedge}\text{-}21\, x^{2.5}\right)\right) / x^{3.40881999999999996} \;/.\; K \to 2.03429473170842056{}^{\wedge}63$$

The complete radial velocity expression is:

$$\left(5.86816742763307708{}^{\wedge}33 \left(-\sqrt{x} + 4.59592600000000039{}^{\wedge}\text{-}15\, x^2 - 4.92270000000000074{}^{\wedge}\text{-}21\, x^{2.5}\right)\right) / x^{3.40881999999999996}$$

    8.3

To test the correctness of the expression, the reciporocal of the radial velocity expression's definite Integral between $a_{G1}$ and present a is calculated:

$$\text{NIntegrate}\left[\left(1 \div \left(\left(5.86816742763307708{}^{\wedge}33 \left(-\sqrt{x} + 4.59592600000000039{}^{\wedge}\text{-}15\, x^2 - 4.92270000000000074{}^{\wedge}\text{-}21\, x^{2.5}\right)\right) / x^{3.40881999999999996}\right)\right), \{x, 3.78579 \times 10^9, 778.3 \times 10^9\}\right]$$

    8.4

$4.56000340850643404{}^{\wedge}9$ years.

The age of Jupiter comes out as assumed. It is 4.56Gy.

Equation 8.4 is used to determine the orbital evolution at the given time intervals. Table 6 gives the orbital evolution of Jupiter.

Table 6. Orbital Evolution of Jupiter.

| Time Before Present(B.P). | Time after the formation of Jupiter. | Jupiter(m) |
|---|---|---|
| 4.56Gy | 0 | 3.78E+09 |
| | 5My | 8.81E+10 |
| | 10My | 1.15E+11 |
| 4.46Gy | 100M | 2.69E+11 |
| | 105M | 2.74E+11 |
| | 110My | 2.78E+11 |
| 4.36GY | 200M | 3.42E+11 |
| | 205M | 3.44E+11 |
| | 210M | 3.47E+11 |
| 4.26Gy | 300M | 3.90E+11 |
| | 305M | 3.92E+11 |
| | 310M | 3.94E+11 |
| 4.16Gy | 400M | 4.27E+11 |
| | 405M | 4.29E+11 |
| | 410M | 4.30E+11 |
| 4.06Gy | 500M | 4.57E+11 |
| | 505M | 4.59E+11 |
| | 510M | 4.60E+11 |
| 3.96Gy | 600M | 4.83E+11 |
| | 605M | 4.84E+11 |
| | 610M | 4.85E+11 |
| 3.86Gy | 700M | 5.05E+11 |
| | 705M | 5.06E+11 |
| | 710M | 5.07E+11 |
| 3.76Gy | 800M | 5.25E+11 |
| | 805M | 5.26E+11 |
| | 810M | 5.27E+11 |
| 3.66Gy | 900M | 5.43E+11 |
| | 905M | 5.43E+11 |
| | 910M | 5.44E+11 |
| 3.56Gy | 1G | 5.58E+11 |
| | 1.005G | 5.59E+11 |
| | 1.010G | 5.60E+11 |

Similarly the radial velocity expression and the corresponding time integral equation is determined and used for calculating Saturn's orbital evolution at intervals of 100My . The same is repeated for different ages of Saturn.

**5. CALCULATION OF SATURN'S ORBITAL EVOLUTION.**

For calculating the Saturn's Orbital Evolution, we calculate the geosynchronous orbits and gravitational resonance point. Next we calculate the structure factors in the expression for radial velocity and we check if the integration of the reciprocal of radial velocity from $a_{G1}$ to present semi-major axis gives the transit time of 4.555Gy as assumed in this paper. It has been assumed that Saturn is formed 5My later or 10My later or 15My or 20My. All these scenarios will be determined to examine the instant of 1:2MMR crossing.

The ω/Ω equation, which is equivalent to lom/lod in planetary satellite dynamics, is set up:

$$E \times x^{1.5} - F \times x^2 \;/.\; \{E \to 20.15789 \times 10^{\wedge}(-16), F \to 1.475 \times 10^{\wedge}(-21)\}$$

$$2.015788999999999983\text{`}^{\wedge}\text{-}15\, x^{1.5`} - 1.47500000000000008\text{`}^{\wedge}\text{-}21\, x^2$$

8.5

Eq. 8.5 is ω/Ω equation and it is tested at the present semi-major axis to see if the calculated value of ω/Ω corresponds to the observed value.

$$2.015788999999999983\text{`}^{\wedge}\text{-}15\, x^{1.5`} - 1.47500000000000008\text{`}^{\wedge}\text{-}21\, x^2 \;/.\; x \to 1427 \times 10^9$$

$432.635273982453316\text{`}$ = calculated value of ω/Ω;

The observed value of ω/Ω = 432.139. Hence Eq.8.5 is correctly set up.

Geo-synchronous orbits radius is calculated by equating Eq.8.5 to UNITY.

$$\text{Solve}[2.015788999999999983\text{`}^{\wedge}\text{-}15\, x^{1.5`} - 1.47500000000000008\text{`}^{\wedge}\text{-}21\, x^2 == 1, x]$$

$\{\{x \to 6.52647141287149246\text{`}^{\wedge}9\}, \{x \to 1.86696927911039303\text{`}^{\wedge}12\}\}$

The above two values are $a_{G1}$ and $a_{G2}$.

Gravitational Resonance point is calculated by equating Eq.8.5 to TWO.

$$\text{Solve}[2.015788999999999983\text{`}^{\wedge}\text{-}15\, x^{1.5`} - 1.47500000000000008\text{`}^{\wedge}\text{-}21\, x^2 == 2, x]$$

$\{\{x \to 1.04776529167135223\text{`}^{\wedge}10\}, \{x \to 1.86624222648434638\text{`}^{\wedge}12\}\}$

Hence $x_2$ ( the point of velocity maxima) is $10.477653 \times 10^9$ m.

Now the exponent M of the structure factor is calculated by equating the first time derivative of radial velocity to ZERO at $x_2$.

$$(-0.5 + M)/x^{0.5} + E \times (2 - M)\, x - F \times (2.5 - M)\, x^{1.5} \;/.$$
$$\{E \to 20.15789 \times 10^{\wedge}(-16), F \to 1.475 \times 10^{\wedge}(-21)\}$$

First E and F numerical values are substituted in the first time derivative of radial velocity. The result is:

$$\frac{-0.5` + M}{x^{0.5`}} + 2.01578899999999983`{}^\wedge\!-15\,(2 - M)\,x - 1.47500000000000008`{}^\wedge\!-21\,(2.5` - M)\,x^{1.5`}$$

Now $x_2$ numerical value is substituted to obtain the first derivative at velocity maxima point:

$$\frac{-0.5` + M}{x^{0.5`}} + 2.01578899999999983`{}^\wedge\!-15\,(2 - M)\,x - 1.47500000000000008`{}^\wedge\!-21\,(2.5` - M)\,x^{1.5`}\ /.\ x \to 1.047765291671 35223`{}^\wedge\!10$$

The final first derivative at $x_2$ is:

$$0.0000211207374953290383`\,(2 - M) - 1.58193280467619734`{}^\wedge\!-6\,(2.5` - M) + 9.76940234532642115`{}^\wedge\!-6\,(-0.5` + M)$$
          8.6

Eq.8.6 is equated to ZERO to obtain the exponent M:

Solve[$0.0000211207374953290383`\,(2 - M) - 1.58193280467619734`{}^\wedge\!-6\,(2.5` - M) + 9.76940234532642115`{}^\wedge\!-6\,(-0.5` + M) == 0$, M]

The root is:

M= {{M → 3.41903635715070297`}}

Now constant K is determined. Through several iterations it is found that $V_{max}$= 278027m/year gives a transit time of 4.555Gy from $a_{G1}$ to present semi-major axis.

Following Mathematica Commands are executed to determine K.
The radial velocity expression is first expressed in terms of m/year by multiplying the m/sec expression by $31.5569088 \times 10^6$ seconds/solar year and substituting the numerical values of E and F the constants of ω/Ω Equation.

$(K \div x\wedge M) \times (E \times x\wedge 2 - F \times x\wedge 2.5 - \sqrt{}\,x) \times$
    $(2 \times (1 + m_1/m_0) \div (m_1\,B)) \times 31.5569088 \times 10\wedge 6\ /.$
  $\{E \to 20.15789 \times 10\wedge(-16),\ F \to 1.475 \times 10\wedge(-21)\}$

The result is:

$$\frac{1}{Bm_1}\left(6.31138175999999884\text{`}^\wedge 7\right.$$
$$K x^{-M}\left(-\sqrt{x} + 2.015788999999999983\text{`}^\wedge\text{-}15\, x^2 - \right.$$
$$\left.\left.1.47500000000000008\text{`}^\wedge\text{-}21\, x^{2.5\text{`}}\right)\right.$$
$$\left.\left(1+\frac{m_1}{m_0}\right)\right)$$

Next mass of Sun $m_0$ and mass of planet Saturn are substituted the radial velocity expression:

$$\frac{1}{Bm_1}\left(6.31138175999999884\text{`}^\wedge 7\right.$$
$$K x^{-M}\left(-\sqrt{x} + 2.015788999999999983\text{`}^\wedge\text{-}15\, x^2 - \right.$$
$$\left.\left.1.47500000000000008\text{`}^\wedge\text{-}21\, x^{2.5\text{`}}\right)\left(1+\frac{m_1}{m_0}\right)\right) /.$$
$$\{m_1 \to 5.69\times 10^{26},\, m_0 \to 1.99\times 10^{30}\}$$

The result is:
$$\frac{1}{B}\left(1.10952308807847965\text{`}^\wedge\text{-}19\right.$$
$$K x^{-M}\left(-\sqrt{x} + 2.015788999999999983\text{`}^\wedge\text{-}15\, x^2 - \right.$$
$$\left.\left.1.47499999999999982\text{`}^\wedge\text{-}21\, x^{2.5\text{`}}\right)\right)$$

Next numerical value of B= √[G(M+m)] is substituted in radial velocity expression:

$$\frac{1}{B}\left(1.10952308807847965\text{`}^\wedge\text{-}19\right.$$
$$K x^{-M}\left(-\sqrt{x} + 2.015788999999999983\text{`}^\wedge\text{-}15\, x^2 - \right.$$
$$\left.\left.1.47499999999999982\text{`}^\wedge\text{-}21\, x^{2.5\text{`}}\right)\right) /.$$
$$B \to 1.152262984\times 10^{10}$$

The result is:
$$9.629078634695423 13\text{`}^\wedge\text{-}30$$
$$K x^{-M}\left(-\sqrt{x} + 2.015788999999999983\text{`}^\wedge\text{-}15\, x^2 - \right.$$
$$\left.1.47499999999999982\text{`}^\wedge\text{-}21\, x^{2.5\text{`}}\right)$$

Next the numerical value of exponent M is substituted:

$$K x^{-M} \left(-\sqrt{x} + 2.01578899999999983\text{\textasciigrave}{}^{\wedge}\text{-}15\, x^2 - 1.47499999999999982\text{\textasciigrave}{}^{\wedge}\text{-}21\, x^{2.5}\right) 9.62907863469542313\text{\textasciigrave}{}^{\wedge}\text{-}30 \;/.\; M \to 3.419$$

The result is:

$$\left(9.62907863469542135\text{\textasciigrave}{}^{\wedge}\text{-}30\, K \left(-\sqrt{x} + 2.01578899999999983\text{\textasciigrave}{}^{\wedge}\text{-}15\, x^2 - 1.47499999999999982\text{\textasciigrave}{}^{\wedge}\text{-}21\, x^{2.5}\right)\right) / x^{3.41900000000000003\text{\textasciigrave}}$$

Now we have the radial velocity expression with only K unknown. To determine K we calculate the expression at $x=x_2$ :

$$\left(9.62907863469542135\text{\textasciigrave}{}^{\wedge}\text{-}30\, K \left(-\sqrt{x} + 2.01578899999999983\text{\textasciigrave}{}^{\wedge}\text{-}15\, x^2 - 1.47499999999999982\text{\textasciigrave}{}^{\wedge}\text{-}21\, x^{2.5}\right)\right) / x^{3.41900000000000003\text{\textasciigrave}} \;/.\; x \to 10.479 \times 10^9$$

The result is $V_{max}$ expression with only K unknown:

$$5.42542895557015292\text{\textasciigrave}{}^{\wedge}\text{-}59\, K$$

As already discussed that by several iterations we have found that $V_{max}$ should be 278027m/year in order to achieve an Age of 4.555Gy for Jupiter. Therefore the above expression is equated to 278027m/yr:

$$\text{Solve}[5.42542912314814795\text{\textasciigrave}{}^{\wedge}\text{-}59\, K == 278027,\, K]$$

The unknown K is:

$$\{\{K \to 5.1245163044111921\text{\textasciigrave}{}^{\wedge}63\}\}$$

By substituting the numerical value of K we set up the complete radial velocity expression:

$$\left(9.62907863469542135\text{\textasciigrave}{}^{\wedge}\text{-}30\, K \left(-\sqrt{x} + 2.01578899999999983\text{\textasciigrave}{}^{\wedge}\text{-}15\, x^2 - 1.47499999999999982\text{\textasciigrave}{}^{\wedge}\text{-}21\, x^{2.5}\right)\right) / x^{3.41900000000000003\text{\textasciigrave}} \;/.\; K \to 5.1245163044111921\text{\textasciigrave}{}^{\wedge}63$$

The complete radial velocity expression is:

$$\left(4.93443704599541455 \times 10^{34} \left(-\sqrt{x} + 2.01578899999999983 \times 10^{-15} x^2 - 1.47499999999999982 \times 10^{-21} x^{2.5}\right)\right) / x^{3.41900000000000003}$$

    8.7

To test the correctness of the expression, the reciporocal of the radial velocity expression's definite Integral between $a_{G1}$ and present a is calculated:

$$\text{NIntegrate}\left[\left(1 \div \left(\left(4.93443704599541455 \times 10^{34} \left(-\sqrt{x} + 2.01578899999999983 \times 10^{-15} x^2 - 1.47499999999999982 \times 10^{-21} x^{2.5}\right)\right) / x^{3.41900000000000003}\right)\right), \{x, 6.52647141287149246 \times 10^9, 1427 \times 10^9\}\right]$$

    8.8

$4.55500296355435807 \times 10^9$ years.

    The age of Saturn comes out as assumed. It is 4.555Gy. In this scenario it is assumed that Saturn forms 5My after Jupiter.

    Equation 8.8 is used to determine the the orbital evolution of Saturn at 100My interval. Table 7 gives the orbital evolution of Saturn , of Jupiter and 2:1 MMR Crossing. The 2:1MMR crossing is identified by measuring the null point in the absolute value of [2- Ps/Pj]. As is evident from the Table 7, the null point occurs at 100My.

Table 7. Orbital Evolution of Saturn(Age 4.555Gy) and MMR Crossing

| B.P. | Time | Jupiter(m) | Saturn(m) | Rs/Rj | Ps/Pj | 2-Ps/Pj | \|2-Ps/Pj\| |
|---|---|---|---|---|---|---|---|
| 4.56Gy | 0 | 3.78E+09 | | | | | |
| 4.555Gy | 5My | 8.81E+10 | 6.53E+09 | 7.41E-02 | 0.020164 | 1.979836 | 1.979836 |
| 4.455Gy | 105M | 2.74E+11 | 4.34E+11 | 1.58E+00 | 1.99539 | 0.00461 | 0.00461 |
| 4.355Gy | 205M | 3.44E+11 | 5.56E+11 | 1.62E+00 | 2.05348 | -0.05348 | 0.05348 |
| 4.255Gy | 305M | 3.92E+11 | 6.40E+11 | 1.63E+00 | 2.085279 | -0.08528 | 0.085279 |
| 4.155Gy | 405M | 4.29E+11 | 7.05E+11 | 1.64E+00 | 2.107382 | -0.10738 | 0.107382 |
| 4.055Gy | 505M | 4.59E+11 | 7.59E+11 | 1.65E+00 | 2.12848 | -0.12848 | 0.12848 |
| 3.955Gy | 605M | 4.84E+11 | 8.06E+11 | 1.66E+00 | 2.145661 | -0.14566 | 0.145661 |
| 3.855Gy | 705M | 5.06E+11 | 8.46E+11 | 1.67E+00 | 2.161099 | -0.1611 | 0.161099 |
| 3.755Gy | 805M | 5.26E+11 | 8.83E+11 | 1.68E+00 | 2.175887 | -0.17589 | 0.175887 |
| 3.655Gy | 905M | 5.43E+11 | 9.16E+11 | 1.69E+00 | 2.189191 | -0.18919 | 0.189191 |
| 3.555Gy | 1.005G | 5.59E+11 | 9.46E+11 | 1.69E+00 | 2.201958 | -0.20196 | 0.201958 |

Table 7 has the discrete values of time at which Jupiter and Saturn's orbit's semi-major axis has been calculated. Fourth Column has the ratio of Saturn's semi major axis and Jupiter's major axis. Fifth Column has the ratio of Saturn's Orbital Period and Jupiter's Orbital Period. Sixth column has the absolute value of [2- Ps/Pj]

$$P_S/P_J = [R_S/R_J]^{3/2}$$
8.9

By inspecting Table 7 we find that 1:2 MMR crossing occurs at 105Myears in the Solar History. So we will inspect 1:2MMR crossing for different Birth Dates of Saturn.

In Figure 1. we give the plot of MMR Dip Graph using Mathematica Commnds of ListPlot.

```
q10 = ListPlot[{{105×10^6, 0.00461},
    {205×10^6, 0.05348}, {305×10^6, 0.085279}, {405×10^6, 0.107382},
    {505×10^6, 0.12848}, {605×10^6, 0.145661}, {705×10^6, 0.161099},
    {805×10^6, 0.175887}, {905×10^6, 0.189191}, {1005×10^6, 0.201958}},
    PlotJoined → True, GridLines → Automatic, Frame → True,
    FrameLabel → Abs (2 – Ratio), PlotLabel → {"2:1 MMR DIP"}]
```

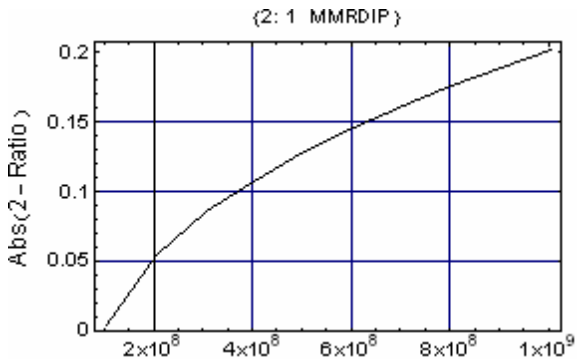

Figure 1.     2:1 MMR Dip graph for Saturn born 5My after Jupiter.

Next Birth date in consideration is 10My after Jupiter's formation. Table 8 gives the evolution of Jupiter and Saturn's Orbital Evolution, Orbital Period Ratio and the absolute value of [2- Ps/Pj]

Table. 8. Orbital Evolution of Saturn(Age 4.55Gy) and MMR Crossing

| B.P. | Time | Jupiter(m) | Saturn(m) | Rs/Rj | Ps/Pj | 2-Ps/Pj | \|2-Ps/Pj\| |
|---|---|---|---|---|---|---|---|
| 4.56Gy | 0 | 3.78E+09 | | | | | |
| | 5My | 8.81E+10 | | | | | |
| 4.55G | 10My | 1.15E+11 | 6.53E+09 | 5.66E-02 | 1.35E-02 | 1.99E+00 | 1.986526 |
| 4.45G | 110My | 2.78E+11 | 4.35E+11 | 1.56E+00 | 1.95E+00 | 5.02E-02 | 0.050241 |
| 4.35G | 210M | 3.47E+11 | 5.56E+11 | 1.60E+00 | 2.03E+00 | -2.82E-02 | 0.028234 |
| 4.25G | 310M | 3.94E+11 | 6.40E+11 | 1.62E+00 | 2.07E+00 | -7.04E-02 | 0.070384 |
| 4.15G | 410M | 4.30E+11 | 7.05E+11 | 1.64E+00 | 2.10E+00 | -9.79E-02 | 0.097868 |
| 4.05G | 510M | 4.60E+11 | 7.59E+11 | 1.65E+00 | 2.12E+00 | -1.21E-01 | 0.12072 |
| 3.95G | 610M | 4.85E+11 | 8.06E+11 | 1.66E+00 | 2.14E+00 | -1.40E-01 | 0.139565 |
| 3.85G | 710M | 5.07E+11 | 8.47E+11 | 1.67E+00 | 2.16E+00 | -1.56E-01 | 0.15624 |
| 3.75G | 810M | 5.27E+11 | 8.83E+11 | 1.68E+00 | 2.17E+00 | -1.71E-01 | 0.171419 |
| 3.65G | 910M | 5.44E+11 | 9.16E+11 | 1.68E+00 | 2.18E+00 | -1.85E-01 | 0.184956 |
| 3.55G | 1.010G | 5.60E+11 | 9.47E+11 | 1.69E+00 | 2.20E+00 | -1.98E-01 | 0.198285 |

By inspection of Table 8 we find that null point of the absolute value of [2- $P_s/P_j$] occurs at 200My. So we next try Saturn born 25My after the formation of Jupiter.

In Figure 2 MMR Dip Graph is given based on Table 8.

```
q20 = ListPlot[{{110×10^6, 0.050241},
    {210×10^6, 0.028234}, {310×10^6, 0.070384}, {410×10^6, 0.097868},
    {510×10^6, 0.12072}, {610×10^6, 0.139565}, {710×10^6, 0.15624},
    {810×10^6, 0.171419}, {910×10^6, 0.184956}, {1010×10^6, 0.198285}},
    PlotJoined → True, GridLines → Automatic, Frame → True,
    FrameLabel → Abs (2 – Ratio), PlotLabel → {"2:1 MMR DIP"}]
```

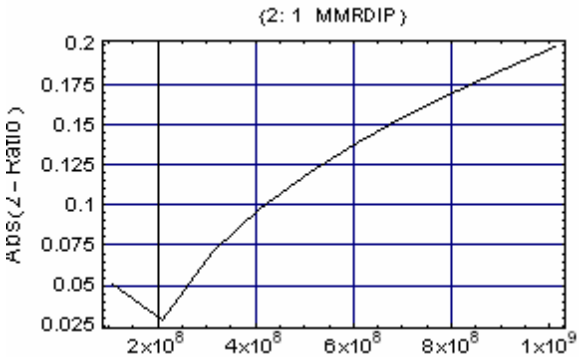

Figure 2.   2:1 MMR Dip graph for Saturn born 10My after Jupiter.

In Table 9 the Orbital Evolution of Jupiter, Saturn , Orbital Periods Ratio and MMR Dip is tabulated.

Table 9. Orbital Evolution of Saturn(Age 4.535Gy) and MMR Crossing

| B.P. | Time | Jupiter(m) | Saturn(m) | Rs/Rj | Ps/Pj | 2-Ps/Pj | \|2-Ps/Pj\| |
|---|---|---|---|---|---|---|---|
| 4.56Gy | 0 | 3.78E+09 | | | | | |
| 4.535G | 25M | 1.63E+11 | 6.53E+09 | 4.00E-02 | 8.01E-03 | 1.99E+00 | 1.991989 |
| 4.435G | 125M | 2.91E+11 | 4.35E+11 | 1.49E+00 | 1.83E+00 | 1.72E-01 | 0.172344 |
| 4.335G | 225M | 3.55E+11 | 5.57E+11 | 1.57E+00 | 1.96E+00 | 3.62E-02 | 0.036236 |
| 4.235G | 325M | 4.00E+11 | 6.41E+11 | 1.60E+00 | 2.03E+00 | -2.70E02 | 0.02699 |
| 4.135G | 425M | 4.35E+11 | 7.06E+11 | 1.62E+00 | 2.07E+00 | -6.68E02 | 0.066753 |
| 4.035G | 525M | 4.64E+11 | 7.60E+11 | 1.64E+00 | 2.10E+00 | -9.63E02 | 0.096252 |
| 3.935G | 625M | 4.89E+11 | 8.07E+11 | 1.65E+00 | 2.12E+00 | -1.20E01 | 0.119779 |
| 3.835G | 725M | 5.10E+11 | 8.48E+11 | 1.66E+00 | 2.14E+00 | -1.41E01 | 0.140662 |
| 3.735G | 825M | 5.30E+11 | 8.84E+11 | 1.67E+00 | 2.16E+00 | -1.58E01 | 0.157512 |
| 3.635G | 925M | 5.47E+11 | 9.17E+11 | 1.68E+00 | 2.17E+00 | -1.74E01 | 0.174016 |
| 3.535G | 1.025G | 5.62E+11 | 9.48E+11 | 1.69E+00 | 2.19E+00 | -1.88E01 | 0.188218 |

In Figure 3. the MMR Dip Graph is plotted for Saturn born 25My after Jupiter.

```
q30 = ListPlot[{{125×10^6, 0.172344},
    {225×10^6, 0.036236}, {325×10^6, 0.02699}, {425×10^6, 0.066753},
    {525×10^6, 0.096252}, {625×10^6, 0.119779}, {725×10^6, 0.140662},
    {825×10^6, 0.157512}, {925×10^6, 0.174016}, {1025×10^6, 0.188218}},
    PlotJoined → True, GridLines → Automatic, Frame → True,
    FrameLabel → Abs (2 − Ratio), PlotLabel → ("2:1 MMR DIP")]
```

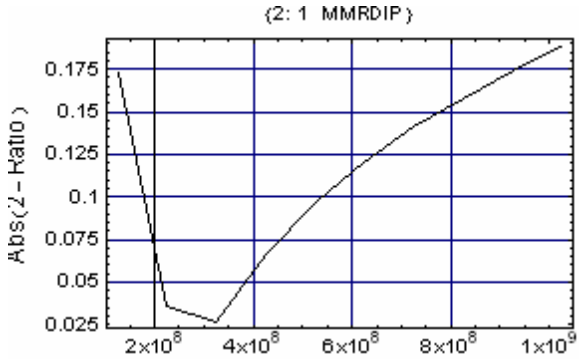

Figure 3.    2:1 MMR Dip graph for Saturn born 25My after Jupiter.

The Kinematic parameters for Jupiter and Saturn are given in Table .10

**Table10. The kinematics parameters of the two Planets.**

|  | Age | $X_2$ ($\times 10^9 m$) | M | K $(N-m^{M+1})$ | $V_{max}$ (m/year) |
|---|---|---|---|---|---|
| **Jupiter** | 4.56Gy | 6.0865 | 3.40884 | **2.0343×10^63** | **203017** |
| **Saturn** | 4.560 |  |  |  |  |
|  | 4.555 | 10.479 | 3.419 | **5.12452×10^63** | **278027** |
|  | 4.550 | 10.479 | 3.419 | **5.13029×10^63** | **278340** |
|  | 4.535 | 10.479 | 3.419 | **514706×10^63** | **279250** |

The superposition of MMR Dip curves are given in Figure 4.
```
Show[{q10, q20, q30}, Axes → True]
```

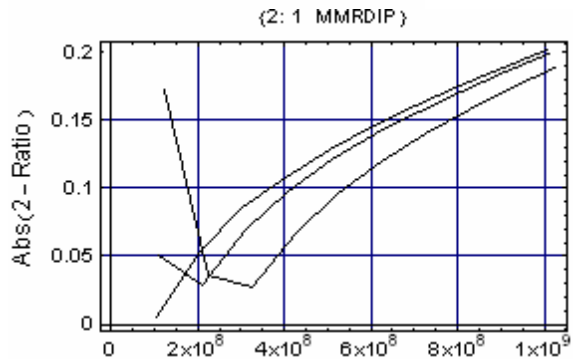

Figure 4. Superposition of MMR Dip Curves for different birth scenarios.

**Table11 . 2:1 MMR crossing dates after the birth of Jupiter.**

| Difference in birth dates(My) | 2:1MMR crossing after the birth of Jupiter in My. |
|---|---|
| 5My | 100My |
| 10My | 200My |
| 25My | 300My |

Late Heavy Bombardment Era [Schoenberg et al 2002] corresponds to 4 Gya to 3.8Gya. This is about 567My to 767My after the birth of Solar Nebula if Solar Nebula is assumed to be born at 4.567Gya. Hence we can safely say that LHB Era occurred from 550My to 750My after the birth of Solar Nebula.. From Table 11. it is evident that we can approach this delay in heavy meteoritic shower only by assuming that Saturn was born 25Myrs later than Jupiter and Jupiter was born at 4.56Gya..

4.   DISCUSSION.

After examining Figure 1, 2, 3 and Table 11 we clearly see that according to our theoretical analysis MMR crossing at best must have taken place about 300My after the birth of Jupiter. At this time 2:1MMR Crossing causes Neptune to be tossed into the outer reaches of Planetesimals disk that surround the entire Solar System called Oort's Cloud [A band of comets believed to be in a spherical shell surrounding our Solar System and extending from 75,000AU to 150,000AU. www.gps.caltech.edy/~mbrown. ]. Neptune gravitationally scatters these planetesimals. This sends a a burst of impactors throughout the entire Solar Sytem resulting into heavy bombardment era.

Time constraints on planetary formation doesnot allow Saturn to be formed any later than 25My after the formation of Jupiter. By 30My the protoplanetary disk of gas and dust gets dissipated. Hence Jupiter, Saturn, Neptune and Uranus must have completed its formation within this narrow time slot. But this permits the 2:1MMR Crossing no later than 300My. But LHB era demands 2:1MMR Crossing in the period of 500My to 700My.

Under the circumstances the only explanation for Late Heavy Bombardment Era is the delay in response to MMR crossing. By further studies only it can be established whether there can be 200My delay in the aftereffects of 2:1 MMR Crossing.

6. CONCLUSIONS.

From Figure 1,2,3 and 4 and from Table 11 we see that 2:1 MMR crossing clearly occurs after 300My from the birth of Jupiter only when birth of Jupiter precedes that of Saturn by 25My.
   This establishes with definiteness that Jupiter was born the earliest probably 4.56Gya and after the birth of this Gas Giant there followed the sequential formation of Saturn, Neptune and Uranus. The Jovian planets formation was completed in 30My after Jupiter. Terrestrial Planets have formed over a much longer time scale of 100My through infrequent Giant Impacts. Venus and Earth was formed in rapid succession followed by Mars and Mercury. By 4.467Gya the Solar System's formation was completed and the last debris in the disc of accretion was completely cleared.
   *Then suddenly what turned on the Late Heavy bombardment era as late as 4.00 Gya*

*and the entire solar system was incessantly battered for almost 200My from 4Gya to 3.8Gya?*

The present paper asserts that Jupiter and Saturn passed through 1:2 MMR. This caused the Neptune to be tossed into the more distant planetismal disc that surrounds the entire Solar System and which is more commonly known as Oort's Cloud. Neptune is flung on an outward migration due to 2:1MMR resonance of Saturn and Jupiter at about 4.267Gya. After migrating to the very edge of our Solar System, Neptune gravitationally scatters this horde of planetismals. This sends a burst of impactors through out the entire system including our Earth and Moon and this is what has come to be known as Late Heavy Bombardment Era and which lasted for almost 200Myr. Because of very large distances involved, the after effects of 2:1MMR resonance was delayed by 267My and its spill over continued for another 200My. This is what has today come to be known as Late Heavy Bombardment Era at 4Gy ago. This heavy bombardment continued for another 200My persisting from 4Gya to 3.8Gya.

This theoretical analysis and its near correspondence with the observed data about the late heavy meteoritic shower clearly corroborates the New Perspective about the birth and evolution of Solar System.

**BIBLOGRAPHY**

| | |
|---|---|
| * | Chaisson, E., McMillan, S. "Astronomy- A beginner's Guide to the Universe" second edition, Prentice Hall, 1998. |
| * | Franklin, Fred A., Lewis, Nikole K., Soper, Paul R. and Holman, Mathew J. , " Hilda Asteroids as Possible Probes of Jovian Migration", *Astronomical Journal,* **128,** 1391, 2004 September. |
| * | Gomes, R., Levison, H.F., Tsiganis, K. and Morbidelli, A. " Origin of the cataclysmic Late Heavy Bombardment period of the terrestrial planets", *Nature*, Vol. **435**, 26 May 2005, pp. 466-469; |
| * | Hannu, K., Kroger, P., Oja, H. et al (Eds.) *Fundamental Astronomy*, Springer, 2003. |
| * | Moore, Sir P., (Ed.) A-Z guide, *Astronomy Encyclopedia*, Oxford Press, 2002. |
| * | Morbidelli, A., Levison, H.F., Tsiganis, K. and Gomes, R. " Chaotic capture of Jupiter's Trojan asteroids in early Solar System", *Nature*, Vol. **435**, 26 May 2005, pp. 462-465; |
| * | Santos, Numo C., Benz W. & Mayor M., "Extra-Solar Planets: Constraints for Planet Formation Model", *Science,* **310,** 251-255, (2005), |
| * | Sharma, B. K. & Ishwar, B. "A New Perspective on the Birth and Evolution of our Solar System based on Planetary Satellite Dynamics", *35th COSPAR Scientific Assembly*, 18-25th July 2004, Paris, France. |
| * | Sharma, B. K. & Ishwar, B., "Jupiter-like Exo-Solar Planets confirm the Migratory Theory of Planets" *Recent Trends in Celestial Mechanics-2004*, pp.225-231, BRA Bihar University, 1st – 3rd Novermber 2004, Muzaffarpur, Bihar.Publisher Elsiever. |
| * | Tsiganis, K., Gomes, R., Morbidelli, A. and Levison, H.F. " Origin of the orbital architecture of the giant planets of our Solar System", *Nature*, Vol. **435**, 26 May 2005, pp. 459-461; |